\documentclass[aps,pre,onecolumn,superscriptaddress,showpacs]{revtex4}
\usepackage[dvips]{graphicx}
\usepackage{amssymb,amsmath}
\usepackage{color}
\usepackage{bm}
\usepackage{float}

\begin{document}
\title{Flexible model of network embedding}
\author{Juan Fern\'andez-Gracia}
\affiliation{Department of Epidemiology, Harvard T.H. Chan School of Public Health, Harvard University}
\affiliation{Instituto de F\'isica Interdisciplinar y Sistemas Complejos IFISC (CSIC - UIB), Palma de Mallorca, E-07122, Spain}
\author{Jukka-Pekka Onnela}
\affiliation{Department of Biostatistics, Harvard T.H. Chan School of Public Health, Harvard University}

\begin{abstract}
There has lately been increased interest in describing complex systems not merely as single networks but rather as collections of networks that are coupled to one another. We introduce an analytically tractable model that enables one to connect two layers in a multilayer network by controlling the locality of coupling. In particular we introduce a tractable model for embedding one network (A) into another (B), focusing on the case where network A has many more nodes than network B. In our model, nodes in network A are assigned, or embedded, to the nodes in network B using an assignment rule where the extent of node localization is controlled by a single parameter. We start by mapping an unassigned ``source'' node in network A to a randomly chosen ``target'' node in network B. We then assign the neighbors of the source node to the neighborhood of the target node using a random walk starting at the target node and with a per-step stopping probability $q$. By varying the parameter $q$, we are able to produce a range of embeddings from local ($q = 1$) to global ($q \to 0$). The simplicity of the model allows us to calculate key quantities, making it a useful starting point for more realistic models.
\end{abstract}

\maketitle

\section*{Introduction}

The last decade has witnessed a transition from the study of simple graphs to more complex structures~\cite{newman2010networks}, such as bipartite networks, temporal networks~\cite{Holme201297}, networks of networks~\cite{interdependent_nets}, and multiplex and multi-layer networks\cite{Kivela01092014}. All these entities have been shown to be of interest for the modeling of different systems at different levels of complexity~\cite{PhysRevE.72.066107,interdependent_applied,PhysRevLett.111.128701}. In the context of multi-layer networks, although there are many generative models for the individual layers that compose them (see for example~\cite{newman2010networks,survey}), the investigation of such models for inter-layer connectivity is scarce and researchers have explored mostly correlations of different centrality measures to establish inter-layer connections, such as by connecting high-degree nodes in one layer to low-degree nodes in the other layer~\cite{buldu2016,byung2018,artime2017}. There are naturally many other ways how nodes in different layers may be coupled, and we introduce a generative and analytically tractable model of inter-layer connectivity that allows for controlling the extent to which locality of connections in one layer is preserved when the network is embedded to the nodes of another layer. We start with a model that consists of pairwise interconnections between layers. We use the term ``embedding'' to refer to an inter-layer connectivity pattern where all nodes of one of the two layers, layer or network A, have at least one link to a node in the other layer, layer or network B), \textit{i.e.}, the coupling is an exhaustive mapping of the nodes of network A to the nodes of network B. Throughout the manuscript, we will use the language of embedding rather than that of multilayer networks. This type of embedding appears naturally in various settings. For example, network A could correspond to a social network of individuals and their pairwise social relationships and network B could represent a network of spatial locations (counties, cities, etc.) and their pairwise connectivity (spatial adjacency, transportation network, etc.).

Note that \emph{embedding} here refers to a different concept than metric embedding of a network into a metric space~\cite{PhysRevLett.100.078701} as we do not aim to have a metric space representation of a network which can substitute the topology of the network and be used, for example, for routing purposes using only local knowledge about the structure of the network~\cite{PhysRevE.82.036106}.

Throughout this paper, we will think of network A as representing a social network and network B a network of geographical locations. Of key interest is the extent of the locality of the embedding. After a ``source'' node in A has been mapped to a ``target'' node in B, locality of the embedding refers to the extent to which the neighbors of the source node in A are mapped to the neighborhood of the target node in B. This characteristic has been studied for social networks embedded in space, finding that the probability of having a social contact at a certain distance is usually well described by a power law with exponent varying between $-0.7$ and $-2$~\cite{Grabowicz2014,Lambiotte20085317,Lengyel2015,Onnela2011}. For this specific setting there are existing models that couple mobility and social connections in order to reproduce specific features of the social networks embedded in geography, such as the similarity of visitation patterns for neighbors in the social network~\cite{Toole2015}, the degree distribution of social networks and the spatial spread of populations~\cite{PhysRevX.4.011008}, or the coupled evolution of mobility and social interactions~\cite{Grabowicz2014}. 

This varying degree of locality has important consequences for processes that take place on a network that is embedded in space. For example, for epidemic processes one can transition from a situation of local embedding, in which spreading fronts move at a well defined velocity over space, to a situation where there are no spreading fronts at all and the spatial description of the epidemics cannot be tackled by a continuous space approach~\cite{Newman2013}. Being able to model this characteristic, the extent of locality of the network embedding, can help one understand emergent properties of dynamical models involving opinion dynamics, epidemic spreading, or diffusion of innovations or cultural traits. 

\section*{The model}

 We consider two undirected and unweighted networks $G_A$ and $G_B$ with $N_A$ and $N_B$ nodes, respectively, and we assume that $N_A > N_B$. Their degree distributions are $P_A(k)$ and $P_B(k)$ and their adjacency matrices are $\bm{A}$ and $\bm{B}$. The assignment rule embeds network A in network B, i.e., each node in $G_A$ will be assigned to a node in $G_B$. Completion of the node assignment process gives rise to a third network called the \emph{embedded network} $G_{\Gamma}$, which is an undirected weighted network with adjacency matrix $\bm{\Gamma}$ consisting of the same same set of nodes as $G_B$. Some of the nodes in $G_A$ will be assigned to node $i$ in $G_B$ and some to node $j$ in $G_B$; the element $\Gamma_{ij}$ of the adjacency matrix of the embedded network corresponds to the number of edges in $G_A$ between the group of nodes assigned to node $i$ in $G_B$ and the group of nodes assigned to node $j$ in $G_B$. We use $f_i$ to denote the \emph{attractiveness} of node $i$ in $G_B$, which is a measure of its ability to attract nodes from network A, and we impose the condition that the attractiveness of all nodes add up to 1 ($\sum_{i=1}^{N_B}f_i=1$) to allow for its interpretation as probability. The values $f_i$ can be assigned based on any rule, which might be coupled to network structure; for example, they could be made proportional to the degrees of nodes in network B or they could depend on an exogenous factor. Returning to the example of a social network A that is embedded in a spatial network B, the attractiveness of nodes in network B could be thought of as the target proportions of agents to embed at each node. Depending on the application, this could reflect the capacities of nodes in network B, which might be proportional to, say, populations of cities. We would like the network embedding process to be carried out such that by varying some coupling parameter, the resulting embedded network $G_{\Gamma}$ has a varying degree of locality, which leads us to the following simple assignment rule which is repeated until all nodes in $G_A$ have been assigned to nodes in $G_B$.

\begin{enumerate}
 \item Choose at random an unassigned node $\alpha$ in $G_A$ (source node) and assign it to node $i$ in $G_B$ (target location) with probability $f_i$.
 \item For each unassigned neighbor of the source node $\alpha$ in $G_A$, start a weighted random walk from the target node $i$ in $G_B$ with a per-step stopping probability $q$ (probability that the random walk stops at each step); assign the node in $G_A$ to the node in $G_B$ where the random walk happens to stop. The random walk on $G_B$ is weighted at each node by the attractiveness of the neighboring (adjacent) nodes.
\end{enumerate}

\begin{figure}
 \centering
 \includegraphics[width=0.5\textwidth]{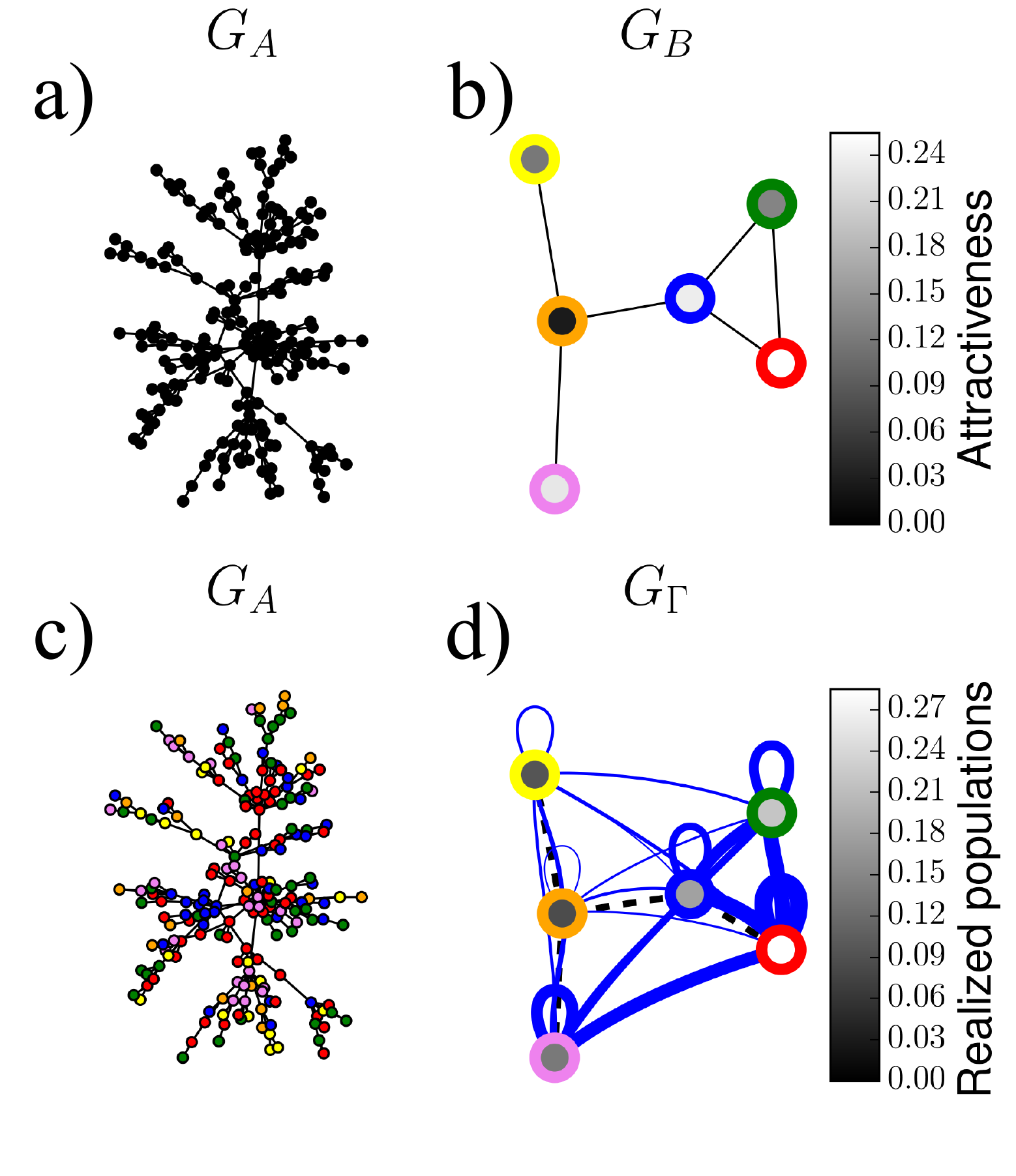}
 \caption{\textbf{Example of the network embedding model.} \textbf{a)} Visualization of network A, here a BA network~\cite{Barabasi509} of size $N_A = 100$ and parameter $m=1$. \textbf{b)} Network B is very simple for illustrative purposes. The outside coloring of each node represents its label and the inside color represents its attractiveness. \textbf{c)} The nodes in A have been assigned to location nodes in B and the color of each network A node matches the outside coloring of a network B node. \textbf{d)} Embedded network $G_{\Gamma}$. The outside coloring of each node is the same as in $G_B$ whereas the inside color represents the percentage of network A nodes that have been assigned to the given node in network B.
 \label{Fig_1}}
\end{figure}

If the stopping probability $q=1$, the walk always (deterministically) comes to a halt at the location node $i$ in $G_B$ where it started, and therefore all unassigned neighbors of a given source node are assigned to the same location as the source node itself. This results in an embedded network $G_{\Gamma}$ that represents the strongest possible coupling between networks A and B. The value of $q=0$ for the stopping probability is not sensible in the context of the proposed model because after assigning a source node, no neighboring nodes would ever be assigned (stopping probability is 0). However, we can consider the limit of $q \to 0$ in which case we simply assign unassigned neighbors of a source node $\alpha$ in $G_A$ to a location node $i$ in $G_B$ with probability that is proportional to the stationary probability of a biased random walker to be at any of the nodes. The probability of being assigned to node $i$ in B will thus be proportional to the corresponding entry of the leading eigenvector $\bm{v}^0$ with eigenvalue 1 of the transition matrix $\bm{C}$ ($C_{ij}=f_iB_{ij}/\sum_kf_kB_{kj}$), which describes the weighted random walk on network B~\cite{serranoVM}.
In this case the embedded network $G_{\Gamma}$ represents the weakest possible coupling between networks A and B. For intermediate values of $0 < q < 1$ the extent of embedding interpolates between these two extreme cases. For a schematic of the model, see Fig.~\ref{Fig_1}.

\section*{Analytical description}

The assignment process stops once every node in $G_A$ has been assigned to a location node in $G_B$. At this point there are several quantities of interest that describe the output of the model, and these include the size of the realized population $\Phi_i$ at each node $i$ in $G_B$, which corresponds to the number of nodes in $G_A$ that have been assigned to location node $i$, and the adjacency matrix $\bm{\Gamma}$ of the embedded network $G_{\Gamma}$. We restrict the analytics to the case where network A has many more nodes than network B, and therefore for network A we make use of the degree distribution $P_A(k)$ only whereas for network B we use the actual adjacency matrix $\bm{B}$. Note that prior to calculating the sizes of the realized populations $\Phi_i$ at location nodes, or the adjacency matrix of the embedded network $\bm{\Gamma}$, we can describe analytically the number of nodes in network $G_A$ that have been assigned to a location node in $G_B$ at different times of the process and how many neighbors they have in $G_A$ with either assigned or unassigned locations in network B. This process can be described independently of any knowledge of the topology of network B or the random walk stopping probability $q$, and since these details are needed for calculating other features of the model, we will start the analytics from there.

\subsection*{Process of location assignment}
\begin{figure}
 \centering
 \includegraphics[width=0.25\textwidth]{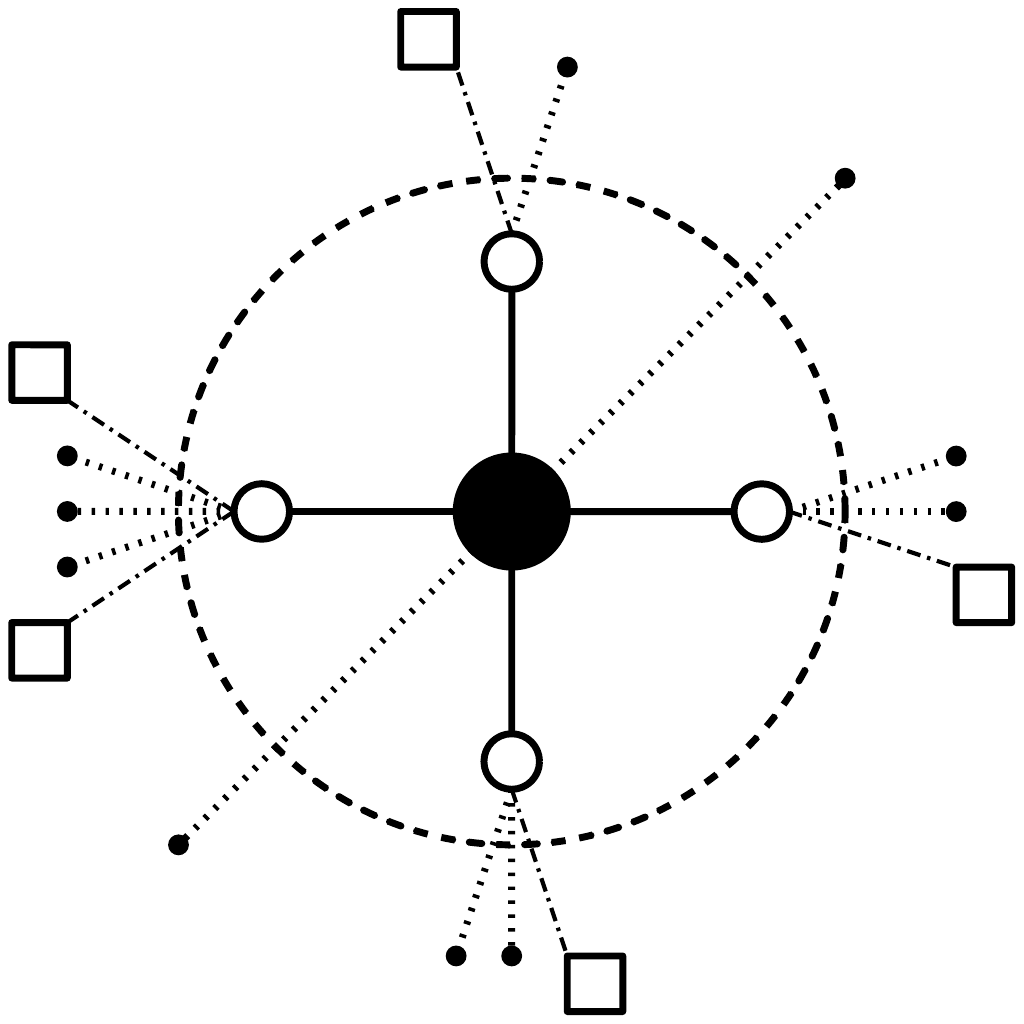}
 \caption{\textbf{Schematic of location assignment.} Shown is a piece of network A that is centered around a node that is chosen at random for location assignment during this time step (shown as the black filled node at the center). The empty circles represent unassigned neighbors of the central node, and the nodes inside the big dashed circle are the ones to be assigned during this time step. The empty squares represent unassigned second neighbors of the central node. Finally, the small filled circles represent assigned nodes. Location assignments in the random walk process will be correlated only for the nodes that are connected with solid edges.
\label{Fig_loc_assignment_q_0}}
\end{figure}

As part of the assignment of nodes at each time step we choose an unassigned source node $\alpha$ in $G_A$ at random and assign the node itself, as well as its unassigned nearest neighbors in $G_A$, to target location nodes in $G_B$. We are interested in two specific quantities related to unassigned nodes in $G_A$ at time $t$, namely, the distribution $P^{\oplus}_A(k,t)$ of the number of their \emph{unassigned} neighbors $k$ (empty circles in Fig.~\ref{Fig_loc_assignment_q_0}) and the distribution $P^{\dagger}_A(k,t)$ of the number of their \emph{assigned} neighbors $k$ (small black circles in Fig.~\ref{Fig_loc_assignment_q_0}). We reduce the description to the computation of the first two moments of $P^{\oplus}_A(k,t)$, $\langle k\rangle_A^{\oplus}$ and $\langle k^2\rangle_A^{\oplus}$, and the first moment of $P^{\dagger}_A(k,t)$, $\langle k\rangle_A^{\dagger}$. Their approximate evolution equations are

\begin{equation}
 \frac{d}{dt}\langle k\rangle^{\oplus}_A(t)=\frac{1}{\eta(t)-1-\langle k\rangle^{\oplus}_A(t)}\left[\langle k\rangle^{\oplus}_A(t)^2-2\langle k^{2}\rangle^{\oplus}_A(t)\right],\label{eqs_loc_assignment_1}
\end{equation}
\begin{equation}
 \frac{d}{dt}\langle k^2\rangle^{\oplus}_A(t)=\frac{1}{\eta(t)-1-\langle k\rangle^{\oplus}_A(t)}
  \times \Bigg\{\langle k^2\rangle^{\oplus}_A(t) \left[1-2\langle k\rangle^{\oplus}_A(t)-2\frac{\langle k^2\rangle^{\oplus}_A(t)}{\langle k\rangle^{\oplus}_A(t)}\right]
  +2\langle k\rangle^{\oplus}_A(t)^3\Bigg\},\label{eqs_loc_assignment_2} 
\end{equation}
\begin{equation}
 \frac{d}{dt}\langle k\rangle^{\dagger}_A(t)=\frac{\langle k^2\rangle^{\oplus}_A(t)}{\eta(t)-1-\langle k\rangle^{\oplus}_A(t)},\label{eqs_loc_assignment_3}
\end{equation}

with initial conditions $\eta(0) = N_A$, $\langle k\rangle^{\oplus}_A(0) = \langle k\rangle_A$, $\langle k^2\rangle^{\oplus}_A(0) = \langle k^2\rangle_A$, and $\langle k\rangle^{\dagger}_A(0) = 0$, 
where $\langle k\rangle_A$ and $\langle k^2\rangle_A$ are the first and second moments of the degree distribution of network A and $\eta(t)$ is the total number of unassigned nodes. The derivation of these equations only assumes that the network is locally tree-like and with no degree-degree correlations, and therefore we would expect it to be reasonably accurate for uncorrelated networks with low levels of clustering. For a complete derivation of the equations see the SI.

After some $t^*$ time steps all nodes in network A will have been assigned, which implies that $N_A=\int_0^{t^*}\left[1+\langle k\rangle^{\oplus}_A(t)\right]dt$.

We can then integrate Eqs. 1--4 together with their initial conditions numerically; here we used the RK4 algorithm but other choices are also possible. 
The agreement is very good between direct simulations from model and numerical integration of the equations for different types of networks of different sizes (see appendix Figs. S2 and S3).

\subsection*{Realized populations} 

\begin{figure}
 \centering
   \includegraphics[width=\textwidth]{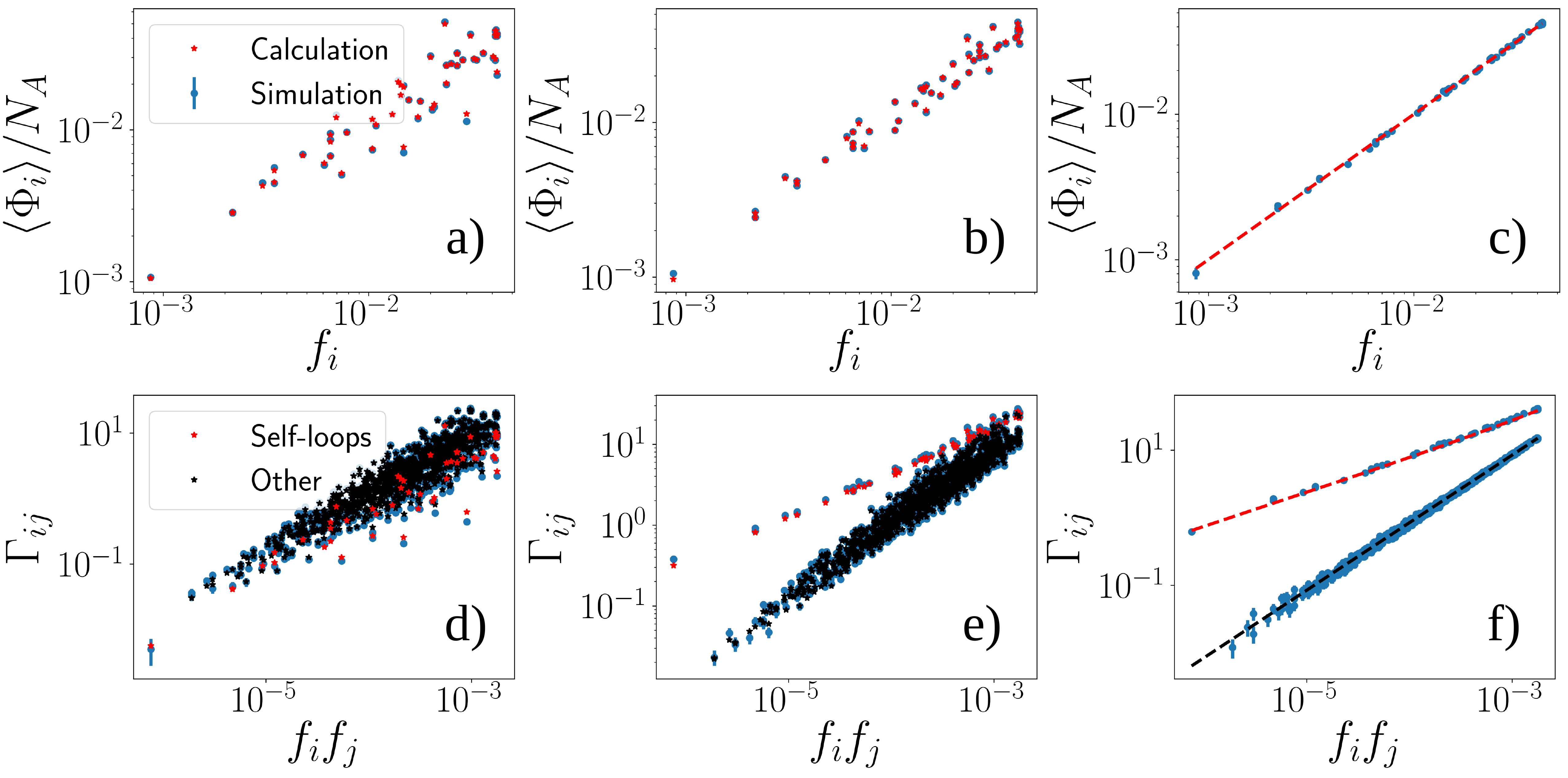}
 \caption{\textbf{Simulation results and analytic results for the outcomes of the model.} $q=0$ (a and d), $q=0.5$ (b and e) and $q=1$ (c and f). Realized populations in the top row and weights in the bottom row. Here $G_A$ corresponds to an Erd\"os-Reny (ER) random graph of size $N_A=10^3$ with edge probability of $p=10^{-2}$. We generate $1000$ network realizations from the ER model and assign them to the target nodes of network B, which here corresponds to a 50-node network that are randomly scattered on a 2-dimensional square and are linked by their spatial adjacency based on a Voronoi tessellation. Results shown are an average over the 1000 independent realizations of the networks. The attractiveness of nodes $f$ in $G_B$ are sampled from a uniform $(1, 100)$ distribution and then rescaled to satisfy the constraint $\sum_{i=1}^{N_B}f_i = 1$. The simulated values are shown in blue while the stars are the results of the calculations. In the bottom row (d, e and f) red stars stand for the weights of the self-loops in the embedded network and black stars for the other edges. Note how the self-loops gain more weight as compared to the rest of edges as $q$ is greater, reflecting the more localized structure of the embedding. In the case $q = 1$ (c and d), the expression for the realized populations and for the weights of the embedded network are simple functions of attractiveness and therefore we show them as curves. \label{Fig_comparing}}
\end{figure}

We use $\Phi_i(t)$ to denote the number of nodes in A that have been assigned to target location node $i$ in B at time $t$. 
In the case $q \neq 0$, the average realized population sizes are given by (see Appendix for derivation):
\begin{equation}
 \langle \Phi_i\rangle=f_i\delta(q)+q\alpha\sum_jf_j\sum_{r=1}^{\infty}(1-q)^r\left[\bm{C}^r\right]_{ij},\label{eq_N_i_sol}
\end{equation}
with $\delta(q)=\int_0^{t^*}\left[1+q\langle k\rangle^{\oplus}_A(t)\right]dt$, $\alpha=\int_0^{t^*}\langle k\rangle^{\oplus}_A(t)dt$ and $C_{ij}=f_iB_{ij}/\sum_lf_lB_{lj}$.
In the case $q=1$, and combining the equation for the realized populations (Eq.~\ref{eq_N_i_sol}), the definition of $\delta(q)$ and the implicit equation for $t^*$, we see that $\langle \Phi_i\rangle(t^*)=N_Af_i$ .
For intermediate values of $q$, the random walk based node assignment process distorts the realized population sizes described by Eq.~\ref{eq_N_i_sol}. For the case $q=0$, the average realized populations will be
\begin{equation}
 \langle \Phi_i\rangle=f_i(N_A-\alpha)+\alpha v_i^0.
\end{equation}
Intuitively, $\alpha$ is the number of nodes that have been embedded through the random walk process. For a more detailed derivation, see the SI. See Fig.~\ref{Fig_comparing} (top row) for a comparison of the analytical solution and simulations for different values of $q$.

\subsection*{Embedded network} 

We now turn to analytical description of the embedded network $G_{\Gamma}$. We investigate average quantities of interest for the adjacency matrix $\bm{\Gamma}$ and the dependence of these quantities on $G_A$, $G_B$, and attractiveness $f_i$ in network B. In the course of the assignment process, pairs of nodes that are connected by an edge in network A can be assigned to network B either synchronously or asynchronously. When two adjacent nodes in $G_A$ are assigned \emph{synchronously}, their assignments (their respective target nodes in network B) will be correlated through the random walk process. To account for synchronous assignment, we calculate an expected adjacency matrix $\bm{\psi}$. The synchronous assignment happens always between a node in A (primary node, the center node in Fig.~\ref{Fig_loc_assignment_q_0}) that is selected for its assignment to a node in B and its not-yet located neighbors in A (secondary nodes, the big empty circles in Fig.~\ref{Fig_loc_assignment_q_0}). Therefore the edge connecting the two nodes from A will connect the two locations of B to which both nodes are assigned, and this locations will be related to each other by the random walk that the secondary nodes perform. The information about the relation between these two locations in B will be stored in the adjacency matrix $\bm{\psi}$ ($\psi_{ij}$ and represents the number of connections in A between the nodes from A assigned to locations $i$ and $j$ in B only due to the process just described, the solid edges in Fig.~\ref{Fig_loc_assignment_q_0}). 
When two adjacent nodes in $G_A$ are assigned \emph{asynchronously}, their assignments to their respective target nodes will be uncorrelated. To account for asynchronous assignment, we calculate the number of \emph{stubs} $\varphi_i$ that a node $i$ in B attains during the assignment process and we construct an expected adjacency matrix given a random stub pairing ($\frac{\varphi_i(t^*)\varphi_j(t^*)}{\sum_l\varphi_l(t^*)}\left(1-\frac{1}{2}\delta_{ij}\right)$). These stubs correspond to the dotted and dashed-dotted lines in~Fig.\ref{Fig_loc_assignment_q_0}; as the former ones correspond to edges in network A that connect nodes that have been assigned a location at a previous step and the latter ones correspond to unlocated second neighbors of the primary node selected for location assignment. See the S.I. for the complete derivation of the equations. 
In the end the adjacency matrix of the embedded network is described by the sum of both contributions as
\begin{equation}\label{eq_gen_ASG}
 \langle \Gamma_{ij}\rangle=\psi_{ij}(t^*)+\frac{\varphi_i(t^*)\varphi_j(t^*)}{\sum_l\varphi_l(t^*)}\left(1-\frac{1}{2}\delta_{ij}\right).
\end{equation}
For the case $q\neq 0$, for which the full derivation is given in the SI, the resulting adjacency matrix of the embedded network is given by Eq.~\ref{eq_gen_ASG} with
\begin{align}
 \psi_{ij}(t^*)&=q\alpha f_i\delta_{ij}+q\alpha\left\{\omega_{ij}+\omega_{ji}\right\}\left(1-\frac{1}{2}\delta_{ij}\right),\\
 \varphi_i(t^*)&=\left(q\beta+\gamma\right)f_i+q\beta\sum_lf_l\left[\bm{\Omega}(q)\right]_{il},
\end{align}
where 
\begin{eqnarray}
 \omega_{ij}=f_j\left[\bm{\Omega}(q)\right]_{ij} \nonumber \\
 \left[\bm{\Omega}(q)\right]_{ij}=\sum_{r=1}^{\infty}(1-q)^r\left[\bm{C}^r\right]_{ij} \nonumber \\
 \alpha=\int_0^{t^*}\langle k\rangle^{\oplus}_A(t)dt \nonumber \\
 \beta=\int_0^{t^*}\left[\langle k^2\rangle^{\oplus}_A(t)-\langle k\rangle^{\oplus}_A(t)+\langle k\rangle^{\oplus}_A(t)\langle k\rangle^{\dagger}_A(t)\right]dt \nonumber \\
 \gamma=\int_0^{t^*}\langle k\rangle^{\dagger}_A(t)dt=N\langle k\rangle_A-2\alpha-\beta. \nonumber
\end{eqnarray}

For the case $q=1$, the matrix $\bm{\Omega}(q)$ vanishes. This matrix is the only element involving the topological information of network $B$. Thus in this case the embedded connections between different nodes in $B$ are random as in the case where $q=0$, but the difference is that now there are many more connections inside each target node. So for $q=1$, putting everything together, Eq.~\ref{eq_gen_ASG} can be written as
\begin{equation}
 \langle \Gamma_{ij}\rangle=f_i\alpha\delta_{ij}+f_if_j\left(N_A\langle k\rangle_A-2\alpha\right)\left(1-\frac{1}{2}\delta_{ij}\right),
\end{equation}
therefore depending only on the attractiveness of network $B$ nodes and the average degree of network $A$.

For the case $q = 0$, the embedded network is given by Eq.~\ref{eq_gen_ASG} with
\begin{align}
 \psi_{ij}(t^*)&=\alpha\left(f_iv_j^0+f_jv_i^0\right)\left(1-\frac{1}{2}\delta_{ij}\right),\\
 \varphi_i(t^*)&=\gamma f_i+\beta v_i^0,
\end{align}
where $v_i^0$ is the $i$'th component of the leading eigenvector of the matrix $\bm{C}$ normalized using the $L_1$-norm. See the SI for more details.

\section*{Discussion}

We have introduced a simple, tractable model of network embedding that regulates the extent of localization of the mapping with a single parameter $q$. The model can also be interpreted within the multilayer network framework, in which case networks A and B correspond to different layers and the assignment of source nodes in network A to target nodes in network B to specifying the interlayer connectivity structure~\cite{Kivela01092014}. We leave it as a future challenge to investigate likely biases in the model that are now induced by the fact that, at each step, two different types of location assignments are done: a source node from network A is assigned randomly to a target node in network B, while the unassigned neighbors of the source node are assigned to the neighborhood of the target location in B via a weighted random walk. For example, choosing a node from network A for location assignment in proportion to its degree may have nontrivial effects on the outcome of the assignment. We would like to stress that the simplicity of the model is what allows us to calculate some of its properties analytically, and as such it can help lay a foundation for the understanding dynamics that take place on embedded networks. We postulate our model thus as a tool for the theoretical exploration for example of epidemic dynamics on social networks that are embedded in geography and so gain a deeper understanding of spatial dynamics derived from purely network dynamics. Consider a situation where we have accurate representations of a social network and a spatial network, where the former is embedded in the latter, but no details are available on how the networks are inter-connected. In this case our model could be used as a null model for the embedding process based on values of the coupling parameter from settings where it is available or can be estimated from data. As a sensitivity analysis, one could then vary the value of this parameter to explore different degrees of locality. Finally as an intralayer connectivity model in the multilayer paradigm this model is also a useful tool in exploring the impact of the locality of the embedding on different settings.

\begin{acknowledgments}
The authors would like to thank Caroline O. Buckee for helpful discussions.
\end{acknowledgments}

\clearpage

\onecolumngrid

\appendix
\renewcommand\thefigure{\thesection\arabic{figure}}  
\setcounter{figure}{0}

\section{Step by step derivation of the model equations}

\section{Step by step derivation of the model equations}

\subsection{Location assignment process}

\begin{figure}
 \centering
 \includegraphics[width=0.45\textwidth]{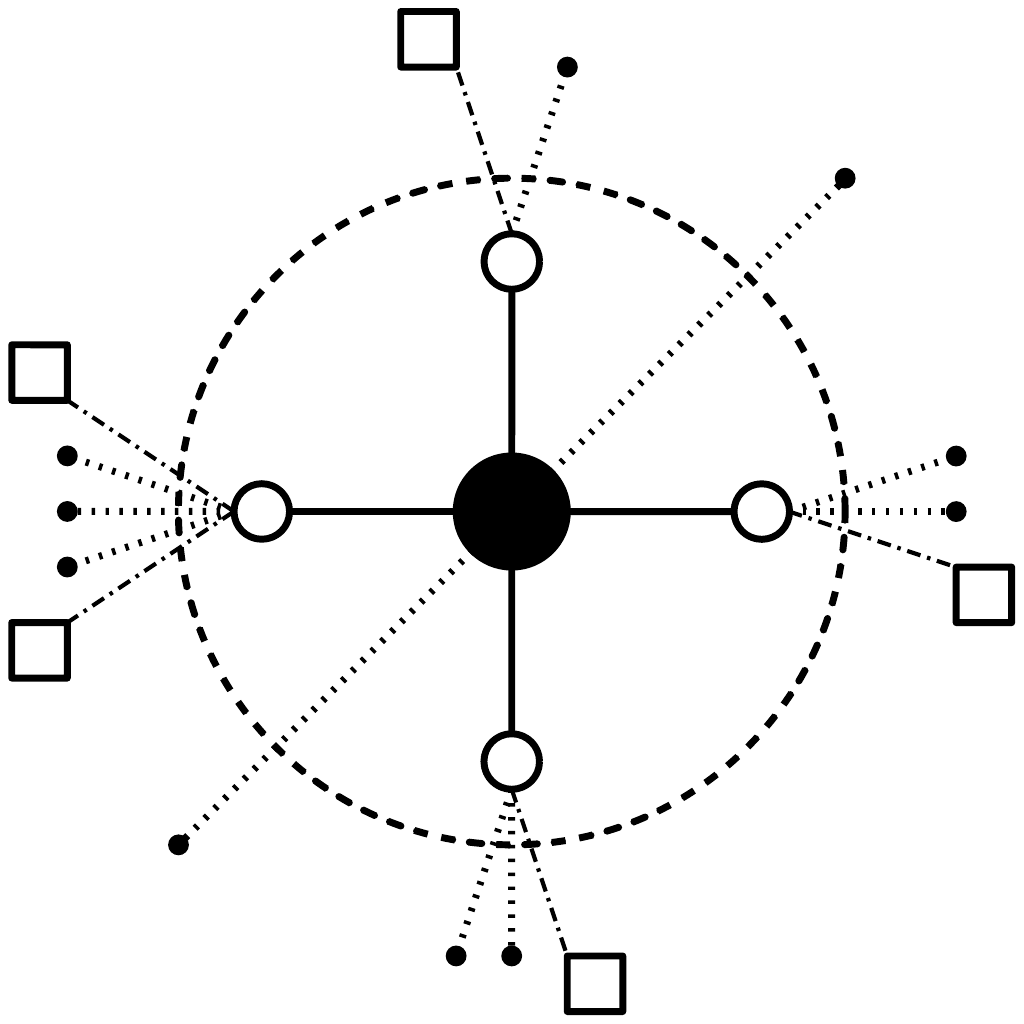}
 \caption{\textbf{Schematic of the process of location assignment.} Shown is a portion of the social network around the node which is chosen for location assignment at this time step. Only nodes up to the second neighborhood are shown. The black filled node in the middle represents the randomly chosen node for location assignment at a certain intermediate time step of the dynamics of location assignment. The empty circles represent the first neighbors of the central node which do not have an assigned location. The nodes inside the large dashed circle are the ones that will be assigned a location in this time step. The empty squares represent second neighbors of the central node with unassigned locations. Finally the small filled circles represent nodes that already have locations assigned.
\label{Fig_loc_assignment_q_0_App}}
\end{figure}

If the stopping probability $q$ equals zero, we assign \emph{location} in $G_B$ to one node from $G_A$. If the parameter $q$ differs from zero, we assign \emph{location} in $G_B$ to a random node and all of its neighbors with unassigned location in $G_A$. The description of the dynamics of this process will help us later for computing several quantities of the embedded network $G_{\Gamma}$. For a node in $G_A$ to be assigned a \emph{location} in $G_B$ is independent of the assigned locations. At each (time) step of the process, we choose a random node from $G_A$ that has not yet been assigned a location and will assign a location in $G_B$ to the node together with all of its first (nearest) neighbors that do not yet have a location assigned. There are two distributions of interest: the distribution $P^{\oplus}_A(k,t)$ of the number of neighbors $k$ of a node in $G_A$ with unassigned location (empty circles in Fig.~\ref{Fig_loc_assignment_q_0_App}) for the nodes in $G_A$ with unassigned location at a certain time step $t$, and the distribution $P^{\dagger}_A(k,t)$ of the number of neighbors $k$ with assigned location (black small circles in Fig.~\ref{Fig_loc_assignment_q_0_App}) for the nodes in $G_A$ with unassigned location at a certain time step $t$.

We first describe the dynamics of $P^{\oplus}_A(k,t)$, \emph{i.e.}, the probability that an unlocated node has $k$ unlocated neighbors at time $t$. We compute the dynamics of the number $\eta^{\oplus}_k(t)$ of unassigned nodes with $k$ unlocated neighbors at time $t$ and from that derive the equations for $P^{\oplus}_A(k,t)=\eta^{\oplus}_k(t)/\eta(t)$, where $\eta(t)=\sum_k\eta^{\oplus}_k(t)$ is the number of unlocated nodes at time $t$. The evolution of $\eta^{\oplus}_k(t)$ depends on three processes
\begin{enumerate}
 \item The node chosen at random for location assignment, from now on called the \emph{central node}, has $k$ unlocated neighbors which happens with probability $P^{\oplus}_A(k,t)$ and reduces the number of nodes in the $k$ category $\eta^{\oplus}_k$ by $1$.
 \item Some neighbors of the central node have $k$ unlocated neighbors. If the central node has $k_i$ unlocated neighbors, the probability of randomly choosing it is $P^{\oplus}_A(k_i,t)$, and the number of unlocated nodes with $k$ unlocated neighbors that are neighbors of the central node is $k_iP^{\oplus}_A(k|k_i)$ (number by which $\eta^{\oplus}_k$ is reduced). $P_A(k|k_i)$ is the probability that an unlocated node with $k$ unlocated neighbors is attached to the central node given that it has $k_i$ unlocated neighbors. Finally we have to sum over all possible values of $k_i$.
 \item A transition process. The second neighbors of the central node loose at least one link to an unlocated neighbor. Here we assume that the network is locally tree-like, so each second neighbor of the central node looses exactly one link. So this process will contribute in two ways: there is a loss for $\eta^{\oplus}_k$ if the second neighbor has $k$ unlocated neighbors, while there is a gain if it has $k+1$ unlocated neighbors. We choose a central node of degree $k_i$ with probability $P_A(k_i,t)$, the number of neighbors of degree $k_j$ is $k_iP^{\oplus}_A(k_j|k_i)$, and finally the number of secondary neighbors of degree $k$ ($k+1$) is $k_jP^{\oplus}_A(k|k_j)$ ($k_jP^{\oplus}_A(k+1|k_j)$) represent a loss (gain) for $\eta^{\oplus}_k$. We have to sum over all possibilities of $k_i$ and $k_j$. 
\end{enumerate}
The conditional probabilities also depend on time, but for clarity we have suppressed the temporal dependence in our notation. Putting everything together we have
\begin{equation}\label{eq_N_k}
 \eta^{\oplus}_k(t+1)=\eta^{\oplus}_k(t)-P^{\oplus}_A(k,t)-\sum_{k_i}P^{\oplus}_A(k_i,t)k_iP^{\oplus}_A(k|k_i)-
 \sum_{k_i}P^{\oplus}_A(k_i,t)k_i\sum_{k_j}P^{\oplus}_A(k_j|k_i)k_j\left[P^{\oplus}_A(k|k_j)-P^{\oplus}_A(k+1|k_j)\right].
\end{equation}
By summing for all $k$ we can see that the total number of unlocated nodes $\eta(t)=\sum_k\eta^{\oplus}_k(t)$  follows
\begin{equation}\label{eq_N}
 \eta(t+1)=\eta(t)-[1+\langle k\rangle^{\oplus}_A(t)],
\end{equation}
with $\langle k\rangle^{\oplus}_A(t)=\sum_kkP^{\oplus}_A(k,t)$. For this result we have used the facts that $\sum_kP^{\oplus}_A(k,t)=1$ and $\sum_kP^{\oplus}_A(k|k')=1$, which are basic normalization properties.

We now return to the formulation of time-dependent probabilities by noting that $P^{\oplus}_A(k,t)=\eta^{\oplus}_k(t)/\eta(t)$, \emph{i.e.}, its dynamical equation (map) is given by dividing Eq.(\ref{eq_N_k}) by Eq.(\ref{eq_N}). We now assume that the network is uncorrelated, which translates to the conditional probabilities
\begin{equation}
 P^{\oplus}_A(k|k')=\frac{k}{\langle k\rangle^{\oplus}_A(t)}P^{\oplus}_A(k,t),
\end{equation}
and therefore
\begin{equation}\label{eq_P_S_map}
 P^{\oplus}_A(k,t+1)=\frac{1}{\eta(t)-1-\langle k\rangle^{\oplus}_A(t)}\left\{P^{\oplus}_A(k,t)\left[\eta(t)-1-k\right]+\frac{\langle k^2\rangle^{\oplus}_A(t)}{\langle k\rangle^{\oplus}_A(t)}\left[(k+1)P^{\oplus}_A(k+1,t)-kP^{\oplus}_A(k,t)\right]\right\}.
\end{equation}
Now let us take the limit to continuous time. We consider $P^{\oplus}_A(k,t+1)-P^{\oplus}_A(k,t)\simeq\partial P^{\oplus}_A(k,t)/\partial t$ and end up with
\begin{equation}\label{eq_P_S_cont}
 \frac{\partial}{\partial t} P^{\oplus}_A(k,t)=\frac{1}{\eta(t)-1-\langle k\rangle^{\oplus}_A(t)}\left\{P^{\oplus}_A(k,t)\left[\langle k\rangle^{\oplus}_A(t)-k\right]+\frac{\langle k^2\rangle^{\oplus}_A(t)}{\langle k\rangle^{\oplus}_A(t)}\left[(k+1)P^{\oplus}_A(k+1,t)-kP^{\oplus}_A(k,t)\right]\right\}.
\end{equation}
Note that this equation depends on the first two moments of the distribution $P^{\oplus}_A(k,t)$ and the number of nodes with unassigned locations $\eta(t)$. Also note that the initial condition is actually $P^{\oplus}_A(k,t=0)=P_A(k)$ as at time $t=0$ all nodes are unlocated.\\
The moment of order $m$ is defined as 
\begin{equation}
 \langle k^m\rangle^{\oplus}_A(t)=\sum_{k=0}^{\infty}k^mP^{\oplus}_A(k,t),
\end{equation}
and we can obtain its dynamical equations by multiplying Eq.(\ref{eq_P_S_cont}) by $k^m$ and summing over $k$ from $0$ to $\infty$. Here we apply that sum over $k$. For the term on $(k+1)P^{\oplus}_A(k+1,t)$ we use the fact that
\begin{align}
 \sum_{k=0}^{\infty}k^m(k+1)P^{\oplus}_A(k+1,t)&=\sum_{i=1}^{\infty}(i-1)^miP^{\oplus}_A(i,t)\\
 &=\sum_{i=0}^{\infty}(i-1)^miP^{\oplus}_A(i,t)\\
 &=\langle k(k-1)^m\rangle^{\oplus}_A(t).
\end{align}

Finally we end up with
\begin{equation}
 \frac{d}{dt}\langle k^m\rangle^{\oplus}_A(t)=\frac{1}{\eta(t)-1-\langle k\rangle^{\oplus}_A(t)}
 \left\{\langle k^m\rangle^{\oplus}_A(t)\langle k\rangle^{\oplus}_A(t)-\langle k^{m+1}\rangle^{\oplus}_A(t)+\frac{\langle k^2\rangle^{\oplus}_A(t)}{\langle k\rangle^{\oplus}_A(t)}\left[\langle k(k-1)^m\rangle^{\oplus}_A(t)-\langle k^{m+1}\rangle^{\oplus}_A(t)\right]\right\}.
\end{equation}
Note that this forms an infinite hierarchy of equations, as the equation for the moment of order $m$ depends on itself, all lower order moments and the moment of order $m+1$. In this type of situation a typical approach is to use a moment closure procedure, which approximates higher order moments by lower order ones. Here we have moments $1$ and $2$ independent and close the hierarchy by approximating the moment of order $3$. One possible approach is to set higher order cumulants to zero. Truncating at order two, the third moment is then given by 
\begin{equation} 
\langle k^3\rangle^{\oplus}_S(t)=3\langle k\rangle^{\oplus}_S(t)\langle k^2\rangle^{\oplus}_S(t)-2\langle k\rangle^{\oplus}_S(t)^3.
\end{equation}

By truncating at second order we set the third cumulant $\kappa_3=0$ and thus imply that the distribution is symmetric as the skewness of a distribution is proportional to its third cumulant. A way around this is to approximate the third moment by its function of the first two if the distribution were lognormal, which yields 
\begin{equation}
 \langle k^3\rangle^{\oplus}_S(t)=\left(\frac{\langle k^2\rangle^{\oplus}_S(t)}{\langle k\rangle^{\oplus}_S(t)^2}+2\right)\langle k\rangle^{\oplus}_S(t)^3\left(\frac{\langle k^2\rangle^{\oplus}_S(t)}{\langle k\rangle^{\oplus}_S(t)^2}-1\right)^2+2\langle k\rangle^{\oplus}_S(t)\langle k^2\rangle^{\oplus}_S(t)-\langle k\rangle^{\oplus}_S(t)^3
\end{equation}

A numerical investigation showed that both prescriptions for moment closure give very similar results. Besides that, by passing to the continuum in Eq.(\ref{eq_N}), we have a closed system of three coupled ordinary differential equations, namely for the first two moments of $P^{\oplus}_S(k,t)$ and the number of remaining nodes without location $\eta(t)$.
\begin{align}
 \frac{d}{dt}\eta(t)&=-\left[1+\langle k\rangle^{\oplus}_A(t)\right] ,\label{eqs_loc_assignment_1}\\
 \frac{d}{dt}\langle k\rangle^{\oplus}_A(t)&=\frac{1}{\eta(t)-1-\langle k\rangle^{\oplus}_A(t)}\left[\langle k\rangle^{\oplus}_A(t)^2-2\langle k^{2}\rangle^{\oplus}_A(t)\right],\\
 \frac{d}{dt}\langle k^2\rangle^{\oplus}_A(t)&=\frac{1}{\eta(t)-1-\langle k\rangle_A(t)}\left\{\langle k^2\rangle^{\oplus}_A(t)\langle k\rangle^{\oplus}_A(t)+\frac{\langle k^2\rangle^{\oplus}_A(t)}{\langle k\rangle^{\oplus}_A(t)}\left[\langle k\rangle^{\oplus}_A(t)-2\langle k^2\rangle^{\oplus}_A(t)\right]-\langle k^{3}\rangle^{\oplus}_A(t)\right\}.\label{eqs_loc_assignment_3}
\end{align}
The initial conditions are given by
\begin{equation}
 \eta(t=0)=N_A\quad,\quad \langle k^m\rangle^{\oplus}_A(t=0)=\sum_kk^mP_A(k).
\end{equation}
Note also that after $t^*$ we will have assigned location to all nodes, which means $\eta(t^*)=0$. Then by integrating the equation for $\eta(t)$ between $t=0$ and $t=t^*$
\begin{equation}
 N_A=\int_0^{t^*}\left[1+\langle k\rangle^{\oplus}_A(t)\right]dt.\label{eq_t_star_App}
\end{equation}

For the distribution $P^{\dagger}_A(k,t)$ of unlocated nodes in $G_A$ with $k$ located nodes at time step $t$, we proceed similarly to the previous case, describing the dynamics of the number $\eta^{\dagger}_k(t)$ (rather than than the proportion) and then approximate the distribution by the proportions. Note that $\sum_k\eta^{\dagger}_k(t)=\sum_k\eta^{\oplus}_k(t)=\eta(t)$ as both sums equal the number of unlocated nodes. There are again three processes by which the number $\eta^{\dagger}_k(t)$ changes:
\begin{enumerate}
 \item $\eta^{\dagger}_k(t)$ decreases by one if the central node chosen for location assignment has $k$ located neighbors. We choose a central node with $k$ located neighbors with probability $P^{\dagger}_A(k,t)$.
 \item $\eta^{\dagger}_k(t)$ decreases by the number of unlocated neighbors of the central node that have $k$ located neighbors. With probability $P^{\oplus}_A(k_i,t)$ the central node has $k_i$ unlocated neighbors and all of them have a probability $P^{\dagger}_A(k,t)$ of having $k$ located neighbors. Finally this term will be summed over all possible values of $k_i$.
 \item There is a transition process. The unlocated second neighbors (second order neighbors via unlocated first order neighbors) of the central node gain one located neighbor, therefore the number of those which have $k$ unlocated neighbors will decrease $\eta^{\dagger}_k(t)$, while the ones with $k-1$ unlocated neighbors will increase it. Here we again assume that the network is locally tree-like.
\end{enumerate}

Putting everything together we end up with
\begin{equation}
 \eta^{\dagger}_k(t+1)=\eta^{\dagger}_k(t)-P^{\dagger}_A(k,t)-\sum_{k_i}P^{\oplus}_A(k_i,t)k_iP^{\dagger}_A(k,t)+\sum_{k_i}P^{\oplus}_A(k_i,t)k_i\sum_{k_j}P^{\oplus}_A(k_j|k_i)k_j\left[P^{\dagger}_A(k-1,t)-P^{\dagger}_A(k,t)\right].
\end{equation}
Assuming that the network is uncorrelated, using that $P^{\dagger}_A(k,t)=\eta^{\dagger}_k(t)/\eta(t)$ and passing to the time continuum limit, we get
\begin{equation}
 \frac{\partial}{\partial t}P^{\dagger}_A(k,t)=\frac{\langle k^2\rangle^{\oplus}_A(t)}{\eta(t)-1-\langle k\rangle^{\oplus}_A(t)}\left[P^{\dagger}_A(k-1,t)-P^{\dagger}_A(k,t)\right].
\end{equation}
Here the initial condition is given by $P^{\dagger}_A(k,t=0)=\delta_{k0}$ since at $t=0$ all nodes are unlocated and thus no node has any located neighbor. Again we can find the dynamical equation for the moment of order $m$, $\langle k^m\rangle^{\dagger}_A(t)$, by multiplying the previous equation by $k^m$ and summing over all $k$ which yields
\begin{equation}
 \frac{d}{dt}\langle k^m\rangle^{\dagger}_A(t)=\frac{\langle k^2\rangle^{\oplus}_A(t)}{\eta(t)-1-\langle k\rangle^{\oplus}_A(t)}\left[\langle (k+1)^m\rangle^{\dagger}_A(t)-\langle k^m\rangle^{\dagger}_A(t)\right].
\end{equation}
Note that the equation for the moments forms a closed system of any maximum order as the equation for the moment of order $m$ does not depend on higher order moments. The initial condition translates for the moments in $\langle k^m\rangle^{\dagger}_A(t=0)=0$. In fact, for the calculations later in the text we only need the first moment whose dynamics are described by
\begin{equation}
 \frac{d}{dt}\langle k\rangle^{\dagger}_A(t)=\frac{\langle k^2\rangle^{\oplus}_A(t)}{\eta(t)-1-\langle k\rangle^{\oplus}_A(t)}.\label{eq_loc_assignment_4}
\end{equation}

\begin{figure}
 \centering
 \includegraphics[width=\textwidth]{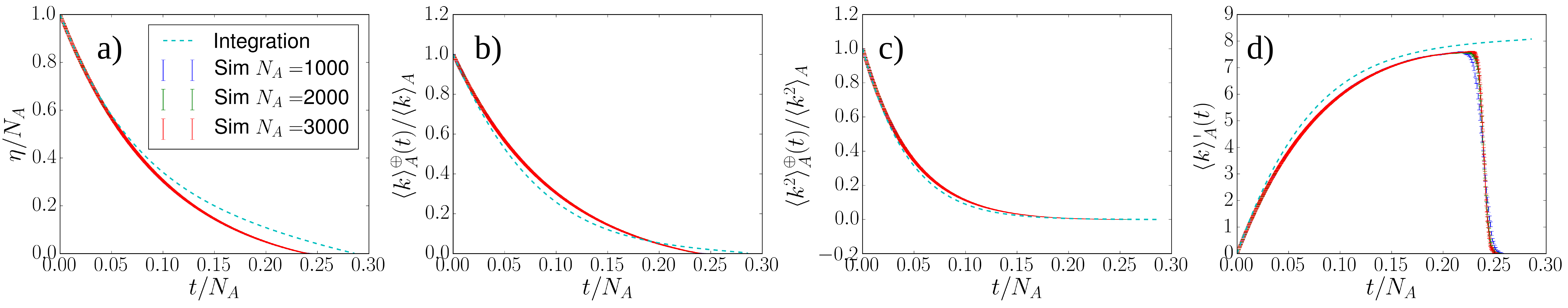}
 \caption{(Color online) \textbf{Assignment results.} We simulated 1000 independent realizations of Erd\"os-Renyi networks of different sizes ($N_A=1000$ in blue, 2000 in green and 3000 in red). Cyan dashed lines for the results of integrating Eqs.~\ref{eqs_loc_assignment_1}--~\ref{eqs_loc_assignment_3} and Eq.~\ref{eq_loc_assignment_4}. The $x$-axis shows the time $t$ normalized by the size of the network, $N_A$. \textbf{a)} Proportion of unlocated nodes $\eta(t)/N_A$. \textbf{b)} Average number of unlocated neighbors for unlocated nodes divided by the average degree of nodes in network A, $\langle k\rangle^{\oplus}_A/\langle k\rangle_A$. \textbf{c)} Second moment of the number of unlocated neighbors for unlocated nodes divided by the second moment of the degree distribution of network A, $\langle k^2\rangle^{\oplus}_A/\langle k^2\rangle_A$. \textbf{d)} Average number of located neighbors for unlocated nodes $\langle k\rangle^{\dagger}_A$.\label{fig_loc_assignment}}
\end{figure}
\begin{figure}
 \centering
 \includegraphics[width=\textwidth]{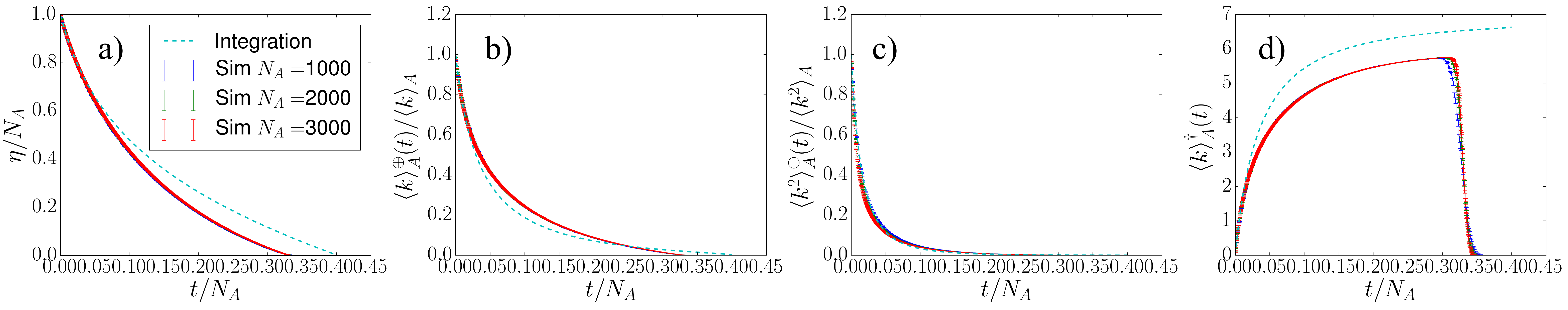}
 \caption{(Color online) \textbf{Location assignment process: agreement between simulations and calculations.} We simulated 1000 independent realizations of Barab\'asi-Albert networks of different sizes ($N_A=1000$ in blue, 2000 in green and 3000 in red). Cyan dashed lines for the results of integrating Eqs.A14--A15 and A21. The $x$-axis shows the time $t$ normalized by the size of the network, $N_A$. \textbf{a)} Fraction of unlocated nodes $\eta(t)/N_A$. \textbf{b)} Average number of unlocated neighbors for unlocated nodes divided by the average degree of nodes in network A, $\langle k\rangle^{\oplus}_A/\langle k\rangle_A$. \textbf{c)} Second moment of the number of unlocated neighbors for unlocated nodes divided by the second moment of the degree distribution of network A, $\langle k^2\rangle^{\oplus}_A/\langle k^2\rangle_A$. \textbf{d)} Average number of located neighbors for unlocated nodes $\langle k\rangle^{\dagger}_A$. \label{fig_loc_assignment_app_BA}}
\end{figure}

\subsection{Realized sizes of node populations}

We use $\Phi_i(t)$ to denote the number of nodes in the social network that have been assigned location $i$. 
In the $q \neq 0$ case, the average realized populations $\langle\Phi_i\rangle(t)$ grow per time step due to two different mechanisms. The first mechanism is that the central node in the location assignment process is assigned location node $i$, what happens with probability $f_i$. The second mechanism is due to unlocated first neighbors of the central node being assigned location $i$. This second mechanism in principle can have infinitely many terms, as many as there are steps in the random walk from $G_A$ to $G_B$. The equation (map) for the average values $\langle \Phi_i\rangle(t)$ is

\begin{align}
 \langle \Phi_i\rangle(t+1)=&\langle \Phi_i\rangle(t)+f_i+f_i\sum_kP^{\oplus}_A(k,t)kq+\sum_jf_j\sum_kP^{\oplus}_A(k,t)k(1-q)\frac{f_iB_{ij}}{\sum_lf_lB_{lj}}q \nonumber\\
 &+\sum_jf_j\sum_kP^{\oplus}_A(k,t)k(1-q)\sum_m\frac{f_mB_{mj}}{\sum_lf_lB_{lj}}(1-q)\frac{f_iB_{im}}{\sum_lf_lB_{lm}}q \nonumber\\
 &+\sum_jf_j\sum_kP^{\oplus}_A(k,t)k(1-q)\sum_m\frac{f_mB_{mj}}{\sum_lf_lB_{lj}}\sum_n(1-q)\frac{f_nB_{nm}}{\sum_lf_lB_{lm}}(1-q)\frac{f_iB_{in}}{\sum_lf_lB_{ln}}q \nonumber\\
 &+\dots \label{eq_N_i}
\end{align}

Here $B_{ij}$ correspond the elements of the adjacency matrix of $G_B$. We define the matrix $C$ with elements 
\begin{equation}
 C_{ij}=\frac{f_i}{\sum_lf_lB_{lj}}B_{ij},
\end{equation}
which specify the probability of a single jump of the random walk starting at location $j$ and ending at location $i$, where the actual jump happens with probability $1-q$). It is straightforward to show that the matrix $C$ has the normalization property $\sum_iC_{ij}=1$, which is the probability of jumping from $j$ to any other location linked in the geographical network. This matrix describes a weighted directed network with the same underlying undirected topology as the geographical network $B$.\\ 
Now using this matrix we can rewrite Eq.~\ref{eq_N_i}, and in the continuous time limit we have
\begin{equation}
 \frac{d}{dt}\langle \Phi_i\rangle(t)=f_i\left[1+q\langle k\rangle^{\oplus}_A(t)\right]+q\langle k\rangle^{\oplus}_A(t)\sum_jf_j\sum_{r=1}^{\infty}(1-q)^r\left[C^r\right]_{ij}, \label{eq_N_i_cont}
\end{equation}
where $\left[C^r\right]_{ij}$ is the element $ij$ of the $r^{\textrm{th}}$ power of  matrix $C$. This element gives the probability of a random walk of distance $r$ from $j$ to $i$ given that the random walk consists of $r$ steps.\\
The initial condition for Eq.~\ref{eq_N_i_cont} is $\langle \Phi_i\rangle(t=0)=0$ because there are no located nodes at $t=0$. After $t^*$ time steps all nodes have been located and the average realized population sizes follow
\begin{equation}
 \langle \Phi_i\rangle(t^*)=f_i\int_0^{t^*}\left[1+q\langle k\rangle^{\oplus}_A(t)\right]dt+q\left(\int_0^{t^*}\langle k\rangle^{\oplus}_A(t)dt\right)\sum_jf_j\sum_{r=1}^{\infty}(1-q)^r\left[C^r\right]_{ij}, \quad (q\ne 0)\label{eq_N_i_sol_App}
\end{equation}
which is obtained by integrating Eq.~\ref{eq_N_i_cont}.\\
In the case $q=1$, combining Eqs.~\ref{eq_N_i_sol_App} and \ref{eq_t_star_App}, we see that $\langle \Phi_i\rangle(t^*)=N_Af_i$ and, as in the $q=0$ case, the realized population sizes are have the same proportions as the input populations.\\
For intermediate values of the stopping probability $q$, the random walk on the network distorts the realized populations as described by Eq.~\ref{eq_N_i_sol_App}.

For the case $q = 0$ we are going to assign first one random source node $\alpha$ of network A to a target node $i$ in network B proportionally to the attractiveness $f_i$ of node $i$. The unassigned neighbors of $\alpha$ will be assigned to a node $j$ in network B proportionally the stationary probability of the weighted random walk. This will be proportional to the corresponding entry of the leading eigenvector $v_j^0$ of the matrix $C$, which is the one encoding the transition probabilities in the random walk. One can also see this from Eq.~\ref{eq_N_i}. In that equation from the third term on on the right hand side, each term describes the random walk of different lengths, stopping at a certain point. Because it is stopping, each term has a factor of $q$, and therefore for $q=0$ they disappear. This is true except for the term at infinity. Therefore the equation is substituted by
\begin{equation}
 \frac{d}{dt}\langle \Phi_i\rangle(t)=f_i+\langle k\rangle^{\oplus}_A(t)\sum_jf_j\lim_{r\to\infty}\left[C^r\right]_{ij}.
\end{equation}
Now we note that the powers of $C$ converge such that $\lim_{r\to\infty}\left[C^r\right]_{ij} = v_i^0$, and as the attractivenesses are normalized, after integrating, we get
\begin{align}
 \langle \Phi_i\rangle(t^*)=&f_it^*+\alpha v_i^0\\
 =&f_i(N_A-\alpha)+\alpha v_i^0, \quad (q\ne 0)
\end{align}
with $\alpha=\int_0^{t^*}\langle k\rangle^{\oplus}_A(t)dt$. Here we are using also that the eigenvector is normalized so that $\sum_i v_i^0 = 1$, \textit{i.e.}, wirh an $L_1$-norm. Similar reasoning as the one presented above will be used to derive further quantities in the case $q = 0$.

\subsection{Embedded network}

Now we turn to the analytical description of the embedded network $G_{\Gamma}$. Again we describe the form of the average for those quantities for the adjacency matrix $\Gamma$, and their dependence on the inputs $G_A$, $G_B$ and attractiveness $f_i$).\\
At each time step, the processes by which two nodes in $G_{\Gamma}$ (this set of nodes is exactly equal to the set of nodes of $G_B$) $i$ and $j$ (which could be the same) gain connections are the following:  
\begin{enumerate}
 \item The links that connect the central node to unlocated neighbors. We can describe this phenomenon through the process of the random walk of the first neighbor and how this process contributes to the number of connections between nodes $i$ and $j$. We use $\psi_{ij}(t)$ to denote the number of connections due to this process at time $t$. These are the solid links in Fig.~\ref{Fig_loc_assignment_q_0_App}.
 \item Node $i$ gains links to an uncorrelated location. For this process, we  count how many \emph{stubs} are accumulated by each node and we approximate how they are linked to other nodes under a random pairing of stubs. We use $\varphi_i(t)$ to denote the total number of these stubs connected to node $i$ at time $t$. This process can happen in two ways ($\varphi_i(t)=\zeta_i(t)+\xi_i(t)$). 
 \begin{enumerate}
  \item Let $\zeta_i(t)$ denote the number of stubs gained by node $i$ up to time $t$ due to neighbors of the central node being located to $i$. These nodes will have links to nodes that are at present unlocated. These are the dashed-dotted links in Fig.~\ref{Fig_loc_assignment_q_0_App}.
  \item Let $\xi_i(t)$ denote the number of stubs gained by location $i$ up to time $t$ due to having nodes located there that have links to already located nodes. These are the dotted links in Fig.~\ref{Fig_loc_assignment_q_0_App}.
 \end{enumerate}
\end{enumerate}
Finally, the average number of social connections will be 
\begin{equation}\label{eq_gen_ASG_App}
 \langle \Gamma_{ij}\rangle=\psi_{ij}(t^*)+\frac{\varphi_i(t^*)\varphi_j(t^*)}{\sum_l\varphi_l(t^*)}\left(1-\frac{1}{2}\delta_{ij}\right)
\end{equation}
In the $q\neq0$ case, the number of correlated connections (solid edges in Fig.~\ref{Fig_loc_assignment_q_0_App}) between nodes $i$ and $j$ will be equal to the probability of locating the central node at $i$ multiplied by the probability of an unlocated neighbor of the central node stopping its random walk at node $j$. We then need to sum the same probability inverting $i$ and $j$, which results in
\begin{align}
 \psi_{ij}(t+1)&=\psi_{ij}(t)+q\sum_kP^{\oplus}_A(k,t)k\left\{f_i\delta_{ij}+\left[f_i\sum_{r=1}^{\infty}(1-q)^r\left[C^r\right]_{ji}+f_j\sum_{r=1}^{\infty}(1-q)^r\left[C^r\right]_{ij}\right]\left(1-\frac{1}{2}\delta_{ij}\right)\right\}\\
 &=\psi_{ij}(t)+q\langle k\rangle^{\oplus}_A(t)\left\{f_i\delta_{ij}+\left[\sum_{r=1}^{\infty}(1-q)^r\left(f_i\left[C^r\right]_{ji}+f_j\left[C^r\right]_{ij}\right)\right]\left(1-\frac{1}{2}\delta_{ij}\right)\right\}.
\end{align}
The term $(1-\frac{1}{2}\delta_{ij})$ corrects for the fact that without it we would be double-counting when $i=j$. Passing to the time continuum we obtain
\begin{equation}
 \frac{d}{dt}\psi_{ij}(t)=q\langle k\rangle^{\oplus}_A(t)\left\{f_i\delta_{ij}+\left[\sum_{r=1}^{\infty}(1-q)^r\left(f_i\left[C^r\right]_{ji}+f_j\left[C^r\right]_{ij}\right)\right]\left(1-\frac{1}{2}\delta_{ij}\right)\right\}.
\end{equation}
Now integrating from $t=0$ to $t=t^*$, with the condition $\psi_{ij}(t=0)=0$, we obtain
\begin{equation}\label{eq_psi}
 \psi_{ij}(t^*)=q\int_0^{t^*}\langle k\rangle^{\oplus}_A(t)dt\left\{f_i\delta_{ij}+\left[\sum_{r=1}^{\infty}(1-q)^r\left(f_i\left[C^r\right]_{ji}+f_j\left[C^r\right]_{ij}\right)\right]\left(1-\frac{1}{2}\delta_{ij}\right)\right\}.
\end{equation}
Now for $\zeta_i(t)$, the number of stubs gained by node $i$ up to time $t$ due to neighbors of the central node being located to node $i$ with unlocated neighbors, we have
\begin{align}
 \zeta_i(t+1)&=\zeta_i(t)+q\left\{f_i\sum_kP^{\oplus}_A(k,t)k\sum_{k'}P^{\oplus}_A(k'|k)
 (k'-1)+\sum_jf_j\sum_kP^{\oplus}_A(k,t)k\sum_{k'}P^{\oplus}_A(k'|k)(k'-1)\sum_{r=1}^{\infty}
 (1-q)^r\left[C^r\right]_{ij}\right\}\\
 &=\zeta_i(t)+q\left[\langle k^2\rangle^{\oplus}_A(t)-\langle k\rangle^{\oplus}_A(t)\right]\left\{f_i+\sum_jf_j\sum_{r=1}^{\infty}(1-q)^r\left[C^r\right]_{ij}\right\}.
 \end{align}
Again, passing to a continuum we have
\begin{equation}
 \frac{d}{dt}\zeta_i(t)=q\left[\langle k^2\rangle^{\oplus}_A(t)-\langle k\rangle^{\oplus}_A(t)\right]\left\{f_i+\sum_jf_j\sum_{r=1}^{\infty}(1-q)^r\left[C^r\right]_{ij}\right\}.
\end{equation}
Now integrating and bearing in mind the initial condition $\zeta_i(t=0)=0$
\begin{equation}\label{eq_zeta}
 \zeta_i(t^*)=q\int_0^{t^*}\left[\langle k^2\rangle^{\oplus}_A(t)-\langle k\rangle^{\oplus}_A(t)\right]dt\left\{f_i+\sum_jf_j\sum_{r=1}^{\infty}(1-q)^r\left[C^r\right]_{ij}\right\}.
\end{equation}
Finally, for $\xi_i(t)$, the number of stubs gained by location $i$ up to time $t$ due to having nodes located there that have links to previously located nodes,  we have
\begin{align}
 \xi_i(t+1)&=\xi_i(t)+f_i\sum_kP^{\dagger}_A(k,t)k+qf_i\sum_kP^{\oplus}_A(k,t)k\sum_{k'}P^{\dagger}_A(k',t)k'+...\\
 &...+q\sum_jf_j\sum_kP^{\oplus}_A(k,t)k\sum_{k'}P^{\dagger}_A(k',t)k'\sum_{r=1}^{\infty}(1-q)^r\left[C^r\right]_{ij}\nonumber\\
 &=\xi_i(t)+f_i\langle k\rangle^{\dagger}_A(t)\left[1+q\langle k\rangle^{\oplus}_A(t)\right]+q\langle k\rangle^{\oplus}_A(t)\langle k\rangle^{\dagger}_A(t)\sum_jf_j\sum_{r=1}^{\infty}(1-q)^r\left[C^r\right]_{ij}.
\end{align}
Treating time as continuous, and integrating from $t=0$ to $t=t^*$ with the initial condition $\xi_i(t=0)=0$, yields
\begin{equation}\label{eq_xi}
 \xi_i(t^*)=f_i\int_0^{t^*}\langle k\rangle^{\dagger}_A(t)dt+q\int_0^{t^*}\langle k\rangle^{\oplus}_A(t)\langle k\rangle^{\dagger}_A(t)dt\left\{f_i+\sum_jf_j\sum_{r=1}^{\infty}(1-q)^r\left[C^r\right]_{ij}\right\}.
\end{equation}
So, summing up, in the general case we have the equation (\ref{eq_gen_ASG_App}) with
\begin{align}
 \psi_{ij}(t^*)&=q\alpha f_i\delta_{ij}+q\alpha\left\{f_i\left[\Omega(q)\right]_{ji}+f_j\left[\Omega(q)\right]_{ij}\right\}\left(1-\frac{1}{2}\delta_{ij}\right),\\
 \varphi_i(t^*)&=\left(q\beta+\gamma\right)f_i+q\beta\sum_lf_l\left[\Omega(q)\right]_{il},\\
 \sum_i\varphi_i&=\beta+\gamma=N_A\langle k\rangle_A-2\alpha,\label{random_stubs_App}
\end{align}
where 
\begin{align}
 \left[\Omega(q)\right]_{ij}&=\sum_{r=1}^{\infty}(1-q)^r\left[C^r\right]_{ij},\\
 \alpha&=\int_0^{t^*}\langle k\rangle^{\oplus}_A(t)dt,\\
 \beta&=\int_0^{t^*}\left[\langle k^2\rangle^{\oplus}_A(t)-\langle k\rangle^{\oplus}_A(t)+\langle k\rangle^{\oplus}_A(t)\langle k\rangle^{\dagger}_A(t)\right]dt,\\
 \gamma&=\int_0^{t^*}\langle k\rangle^{\dagger}_A(t)dt=N\langle k\rangle_A-2\alpha-\beta.
\end{align}
For Eq.(\ref{random_stubs_App}) note that we count all stubs of edges whose ends were not assigned a location during that same time step. 

For the $q=1$ case, note that the last terms inside curly brackets in Eqs.~\ref{eq_psi}, \ref{eq_zeta} and \ref{eq_xi} vanish. Those terms are the only ones involving geographical network information. Thus, in this case, the embedded connections between different locations are random as in the  $q=0$ case. The difference now is that there are many more social connections inside the same location. So for $q=1$, putting everything together, Eq.~\ref{eq_gen_ASG_App} is written as
\begin{equation}
 \langle \Gamma_{ij}\rangle=f_i\alpha\delta_{ij}+f_if_j\left(N_A\langle k\rangle_A-2\alpha\right)\left(1-\frac{1}{2}\delta_{ij}\right), \quad (q=1).
\end{equation}

Following similar reasoning as with the calculation of the realized populations for the case $q = 0$, one can derive that the embedding network in this case will be Eq.~\ref{eq_gen_ASG_App} with
\begin{align}
 \psi_{ij}(t^*)&=\alpha\left(f_iv_j^0+f_jv_i^0\right)\left(1-\frac{1}{2}\delta_{ij}\right),\\
 \varphi_i(t^*)&=\gamma f_i+\beta v_i^0,
\end{align}
where $v_i^0$ is the $i$'th component of leading eigenvector of the matrix C normalized with an $L_1$-norm.

\section{Proportion of correlated embedded connections}

We can compute the fraction $\rho$ of all links of the social network whose end locations are correlated through the random walk:
\begin{equation}
 \rho=\frac{\int_0^{t^*}\langle k\rangle^{\oplus}_A(t)dt}{N_A\langle k\rangle_A}=\frac{\alpha}{N_A\langle k\rangle_A}.
\end{equation}

\section{Possible mechanisms for enhancing the extent of correlations}

The proportion of links with correlated locations can be increased by choosing the central node to be located in proportion to the number of unlocated neighbors. In this case we would have to recalculate the evolution equations and other quantities dealt with above. The proportion of correlated links would now be given by
\begin{equation}
 \rho=\frac{\int_0^{t^*}\frac{\langle k^2\rangle^{\oplus}_A(t)}{\langle k\rangle^{\oplus}_A(t)}dt}{N_A\langle k\rangle_A}.
\end{equation}

Another possibility is to locate not only the first neighborhood of the central node, but also further neighborhoods. We anticipate that this would further complicate the analytical treatment of the quantities of interest.


\begin{thebibliography}{10}

\bibitem{newman2010networks}
M.~Newman.
\newblock {\em Networks: An Introduction}.
\newblock OUP Oxford, 2010.

\bibitem{Holme201297}
P.~Holme and J.~Saram\"{a}ki.
\newblock Temporal networks.
\newblock {\em Physics Reports}, 519(3):97 -- 125, 2012.
\newblock Temporal Networks.

\bibitem{interdependent_nets}
S.G.~Buldyrev, J.~Gao and S.~Havlin.
\newblock Networks formed from interdependent networks.
\newblock {\em Nature Physics}, 8:40 -- 48, 2012.

\bibitem{Kivela01092014}
Mikko Kivelä, Alex Arenas, Marc Barthelemy, James~P. Gleeson, Yamir Moreno,
  and Mason~A. Porter.
\newblock Multilayer networks.
\newblock {\em Journal of Complex Networks}, 2(3):203--271, 2014.

\bibitem{PhysRevE.72.066107}
R.~Lambiotte and M.~Ausloos.
\newblock Uncovering collective listening habits and music genres in bipartite networks.
\newblock {\em Phys. Rev. E}, 72:066107, Dec 2005.

\bibitem{interdependent_applied}
P.~Gerald H.E.~Stanley S.G.~Buldyrev, R.~Parshani and S.~Havlin.
\newblock Catastrophic cascade of failures in interdependent networks.
\newblock {\em Nature}, 464:1025 -- 1028, 2010.

\bibitem{PhysRevLett.111.128701}
C.~Granell, S.~G\'omez, and A.~Arenas.
\newblock Dynamical interplay between awareness and epidemic spreading in
  multiplex networks.
\newblock {\em Phys. Rev. Lett.}, 111:128701, Sep 2013.

\bibitem{survey}
A.~Goldenberg, A.X.~Zheng, S.E.~Fienberg and Ed.M.~Airoldi
\newblock A Survey of Statistical Network Models
\newblock {\em Foundations and Trends in Machine Learning}, Vol. 2: No. 2, pp 129-233 2010.

\bibitem{buldu2016}
J.M.~Buld\'u, R.~Sevilla-Escoboza, J.~Aguirre, D.~Papo D., R.~Gutiérrez
\newblock Interconnecting Networks: The Role of Connector Links.
\newblock In: Garas A. (eds) {\em Interconnected Networks} Understanding Complex Systems. Springer, Cham 2016.

\bibitem{byung2018}
B.~Min, S.D.~Yi, K.-M.~Lee, and K.-I.~Goh
\newblock Network robustness of multiplex networks with interlayer degree correlations
\newblock {\em Phys. Rev. E} 89, 042811 2014.

\bibitem{artime2017}
O.~Artime, J.~Fernández-Gracia, J.J.~Ramasco and M.~San Miguel
\newblock Joint effect of ageing and multilayer structure prevents ordering in the voter model
\newblock {\em Scientific Reports} 7, 7166 2017.

\bibitem{PhysRevLett.100.078701}
M.~\'Angeles Serrano, Dmitri Krioukov, and Mari\'an Bogu\~n\'a.
\newblock Self-similarity of complex networks and hidden metric spaces.
\newblock {\em Phys. Rev. Lett.}, 100:078701, Feb 2008.

\bibitem{PhysRevE.82.036106}
D.~Krioukov, F.~Papadopoulos, M.~Kitsak, A.~Vahdat, and
  M.~Bogu\~n\'a.
\newblock Hyperbolic geometry of complex networks.
\newblock {\em Phys. Rev. E}, 82:036106, Sep 2010.

\bibitem{Grabowicz2014}
P.A.~Grabowicz, J.J.~Ramasco, B.~Gon\c{c}alves, and
  V.M.~Egu\'{\i}luz.
\newblock {Entangling mobility and interactions in social media.}
\newblock {\em PloS one}, 9(3):e92196, January 2014.

\bibitem{Lambiotte20085317}
R.~Lambiotte, V.D.~Blondel, C.~de~Kerchove, E.~Huens,
  C.~Prieur, Z.~Smoreda, and P.~Van Dooren.
\newblock Geographical dispersal of mobile communication networks.
\newblock {\em Physica A: Statistical Mechanics and its Applications},
  387(21):5317 -- 5325, 2008.

\bibitem{Lengyel2015}
B.~Lengyel, A.~Varga, B.~S\'{a}gv\'{a}ri, \'{A}.~Jakobi, and
  J.~Kert\'{e}sz.
\newblock {Geographies of an online social network}.
\newblock page~17, March 2015.

\bibitem{Onnela2011}
J.~P.~Onnela, S.~Arbesman, M.~C.~Gonz\'{a}lez,
  A.~L.~Barab\'{a}si, and N.~A.~Christakis.
\newblock {Geographic constraints on social network groups}.
\newblock {\em PLoS ONE}, 6(4):e16939, January 2011.

\bibitem{Toole2015}
J.~L.~Toole, C.~Herrera-Yag\"{u}e, C.~M.~Schneider, and M.~C.~Gonz\'{a}lez.
\newblock {Coupling human mobility and social ties.}
\newblock {\em Journal of the Royal Society, Interface / the Royal Society},
  12(105):20141128--, April 2015.

\bibitem{PhysRevX.4.011008}
G.~F.~Frasco, J.~Sun, H.~D.~Rozenfeld, and D.~ben~Avraham.
\newblock Spatially distributed social complex networks.
\newblock {\em Phys. Rev. X}, 4:011008, Jan 2014.

\bibitem{Newman2013}
S.~A.~Marvel, T.~Martin, C.~R.~Doering, D.~Lusseau, and M.~E.~J.~Newman.
\newblock The small-world effect is a modern phenomenon.
\newblock arXiv:1310.2636v1 [physics.soc-ph] October 2013.

\bibitem{serranoVM}
M.~A.~Serrano, K.~Klemm, F.~Vazquez, V.~M.~Egu{\'{\i}}luz, and M.~San~Miguel
\newblock Conservation laws for voter-like models on random directed networks
\newblock {\em Journal of Statistical Mechanics: Theory and Experiment}, 2009:10, P10024, Oct 2009

\bibitem{code_url}
{Python code for the model}.
\newblock \url{https://github.com/onnela-lab/network-embedding}.
\newblock Accessed: 2019-12-16.

\bibitem{Barabasi509}
A.-L.~Barab{\'a}si and R.~Albert.
\newblock Emergence of scaling in random networks.
\newblock {\em Science}, 286(5439):509--512, 1999.

\end{thebibliography}
\end{document}